\documentclass[sigconf, 11pt]{acmart}
\renewcommand\footnotetextcopyrightpermission[1]{} 
\usepackage{booktabs} 

\setcopyright{rightsretained}

\acmDOI{}

\acmISBN{}

\acmConference[FATML'17]{ACM FATML Workshop}{August 2017}{Halifax, NS, Canada} 
\acmYear{2017}
\copyrightyear{2017}

\acmPrice{15.00}

\begin{document}
\title{Multisided Fairness for Recommendation}

\author{Robin Burke}
\orcid{0000-0001-5766-6434}
\affiliation{%
  \institution{School of Computing}
  \streetaddress{DePaul University}
  \city{Chicago} 
  \state{Illinois} 
}
\email{rburke@cs.depaul.edu}

\begin{abstract}
Recent work on machine learning has begun to consider issues of fairness. In this paper, we extend the concept of fairness to recommendation. In particular, we show that in some recommendation contexts, fairness may be a multisided concept, in which fair outcomes for multiple individuals need to be considered. Based on these considerations, we present a taxonomy of classes of fairness-aware recommender systems and suggest possible fairness-aware recommendation architectures.
\end{abstract}


\maketitle

\section{Introduction}

\noindent Bias and fairness in machine learning are topics of considerable recent research interest~\cite{pedreshi2008discrimination,fairness,bozdag_bias_2013}. A standard approach in this area is to identify a variable or variables representing membership in a protected class, for example, race in an employment context, and to develop algorithms that remove bias relative to this variable. See, for example, ~\cite{zemel2013learning,Kamishima2012,kamiran2010discrimination,zhang2017anti}.{\let\thefootnote\relax\footnote{Presented as a poster at the 2017 Workshop on Fairness, Accountability, and Transparency in Machine Learning}}

To extend this concept to recommender systems, we must recognize the key role of personalization. Inherent in the idea of recommendation is that the best items for one user may be different than those for another. It is also important to note that recommender systems exist to facilitate transactions. Thus, many recommendation applications involve multiple stakeholders and therefore may give rise to fairness issues for more than one group of participants~\cite{abdollahpouri_recommender_2017}.

\subsection{Personalization}

The dominant recommendation paradigm, collaborative filtering~\cite{koren2011advances}, uses user behavior as its input, ignoring user demographics and item attributes. However, this does not mean that fairness is irrelevant. Consider a recommender system suggesting job opportunities to job seekers. An operator of such a system might wish, for example, to ensure that male and female users with similar qualifications get recommendations of jobs with similar rank and salary. The system would therefore need to defend against biases in recommendation output, even biases that might arise entirely due to behavioral differences: for example, male users might be more likely to click optimistically on high-paying jobs. 

Defeating such biases is difficult if we cannot assert a shared global preference ranking over items. Personal preference is the essence of recommendation especially in areas like music, books, and movies where individual taste is paramount. Even in the employment domain, some users might prefer a somewhat lower-paying job if it had other advantages: such as flexible hours, shorter commute time, or better benefits. Thus, to achieve the policy goal of fair recommendation of jobs by salary, a site operator will have to go beyond a purely personalization-oriented approach, identify salary as the key outcome variable, and control the recommendation algorithm to make it sensitive to the salary distribution for protected groups.

\subsection{Multiple stakeholders}

As the example of job recommendation makes clear, a recommender system is often in the position of facilitating a transaction between parties, such as job seeker and prospective employer. Fairness towards both parties may be important. For example, at the same time that the system is ensuring that male and female users to get recommendations with similar salary distributions, it might also need to ensure that jobs at minority-owned businesses are being recommended to the most desirable job candidates at the same rate as jobs at white-owned businesses.

A \textit{multistakeholder recommender system} is one in which the end user is not the only party whose interests are considered in generating recommendations~\cite{soappaper,abdollahpouri_recommender_2017}. This term acknowledges that recommender systems often serve multiple goals and therefore a purely user-centered approach is insufficient. Bilateral considerations, such as those in employment recommendation, were first studied in the category of \textit{reciprocal recommendation} where a recommendation must be acceptable to both parties in a transaction~\cite{akoglu_valuepick:_2010}. Other reciprocal recommendation domains include on-line dating~\cite{reciprocal}, peer-to-peer ``sharing economy'' recommendation (such as AirBnB, Uber and others), on-line advertising \cite{targetadvertisingbiding}, and scientific collaboration~\cite{lopes2010collaboration,tang2012cross}.

When recommendations must account for the needs of more than just the two transacting parties, we move beyond reciprocal recommendation to multistakeholder recommendation. Today's web economy hosts a profusion of multisided platforms, systems of commerce and exchange that bring together multiple parties in a marketplace, where the transacting individuals and the market itself all share in the transaction~\cite{evans_matchmakers:_2016}. These platforms must by design try to satisfy multiple stakeholders. Examples include LinkedIn, which brings together professionals, employers and recruiters; Etsy, which brings together shoppers and small-scale artisans; and Kiva.org, which brings together charitably-minded individuals with third-world entrepreneurs in need of capital. 

\subsection{Stakeholder utility}

Different recommendation scenarios can be distinguished by differing configurations of interests among the stakeholders. We divide the stakeholders of a given recommender system into three categories: consumers $C$, providers $P$, and platform or system $S$. The consumers are those who receive the recommendations. They are the individuals whose choice or search problems bring them to the platform, and who expect recommendations to satisfy those needs. The providers are those entities that supply or otherwise stand behind the recommended objects, and gain from the consumer's choice.\footnote{In some recommendation scenarios, like on-line dating, the consumers and providers are same individuals.} The final category is the platform itself, which has created the recommender system in order to match consumers with providers and has some means of gaining benefit from successfully doing so. 

Recommendation in multistakeholder settings needs to be approached differently from user-focused environments. In particular, we have found that formalizing and computing stakeholder utilities is a productive way to design and evaluate recommendation algorithms. Ultimately, the system owner is the one whose utility should be maximized: if there is some outcome valued by the recommender system operator, it should be included in the calculation of system utility. 

The system inevitably has objectives that are a function of the utilities of the other stakeholders. Multisided platforms thrive when they can attract and retain critical masses of participants on all sides of the market. In our employment example, if a job seeker does not find the system's recommendations valuable, he or she may ignore this aspect of the system or may migrate to a competing platform. The same is true of providers; a company may choose other platforms on which to promote its job openings if a given site does not present its ads as recommendations or does not deliver acceptable candidates.

System utilities are highly domain-specific: tied to particular business models and types of transactions that they facilitate. If there is some monetary transaction facilitated by the platform, the system will usually get some share. The system will also have some utility associated with customer satisfaction, and some portion of that can be attributed to providing good recommendations. In domains subject to legal regulation, such as employment and housing, there will be value associated with compliance with anti-discrimination statutes. There may also be a (difficult to quantify) utility associated with an organization's social mission that may also value fair outcomes. All of these factors will govern how the platform values the different trade-offs associated with making recommendations.

\section{Multisided fairness}

Recommendation processes within multisided platforms can give rise to questions of multisided fairness. Namely, there may be fairness-related criteria at play on more than one side of a transaction, and therefore the transaction cannot be evaluated simply on the basis of the results that accrue to one side. There are three classes of systems, distinguished by the fairness issues that arise relative to these groups: consumers (C-fairness), providers (P-fairness), and both (CP-fairness).

\subsection{C-fairness}

A recommender system distinguished by C-fairness is one that must take into account the disparate impact of recommendation on protected classes of recommendation consumers. In the motivating example from~\cite{fairness}, a credit card company is recommending consumer credit offers. There are no producer-side fairness issues since the products are all coming from the same bank. 

Multistakeholder considerations do not arise in systems of this type. A number of designs could be proposed. One intriguing possibility is to design a recommender system following the approach of \cite{zemel2013learning} in generating fair classification. We could define a mapping from each user to a prototype space, perhaps defined in terms of latent factors extracted from the rating data. Each prototype could be engineered to have the property of statistical parity relative to the protected class. A key consideration in this type of system is to ensure a bounded loss with respect to ranking accuracy for users. We plan to explore this type of system design in future work. 

\subsection{P-fairness}

A system of requiring P-fairness is one in which fairness needs to be preserved for the providers only. A good example of this kind of system is Kiva.org, on-line micro-finance site. Kiva aggregates loan requests from field partners around the world who lend small amounts of money to entrepreneurs in their local communities. The loans are funded interest-free by Kiva's members, largely in the United States. Kiva does not currently offer a personalized recommendation function, but if it did, one can imagine a goal of the organization would be to preserve fair distribution of capital across its different partners in the face of well-known biases of users~\cite{lee2014fairness}. Consumers of the recommendations are essentially donors and do not receive any direct benefit from the system, so there are no fairness considerations on the consumer side. 

P-fairness may also be a consideration where there is interest in ensuring market diversity and avoiding monopoly domination. For example, in the on-line craft marketplace Etsy\footnote{www.etsy.com}, the system may wish to ensure that new entrants to the market get a reasonable share of recommendations even though they will have had fewer shoppers than established vendors. This type of fairness may not be mandated by law, but is rooted instead in the platform's business model.

There are complexities in P-fairness systems that do not arise in the C-fairness case. In particular, the producers in the P-fairness case are passive; they do not seek out recommendation opportunities but rather must wait for users to come to the system and request recommendations. Consider the employment case discussed above. We would like it to be the case that jobs at minority-owned businesses are recommended to highly-qualified candidates at the same rate that jobs at other types of businesses. The opportunity for a given minority-owned business to be recommended to an appropriate candidate may arrive only rarely and must be recognized as such. As with the C-fairness case, we will want to bound the loss of personalization that accompanies any promotion of protected providers. 

There is considerable research in the area of diversity-aware recommendation~\cite{Vargas:2011:RRN:2043932.2043955,adomavicius2012improving}. Essentially, these systems treat recommendation as a multi-objective optimization problem where the goal is to maintain a certain level of accuracy, while also ensuring that recommendation lists are diverse with respect to some representation of item content. These techniques can be re-purposed for P-fairness recommendation by treating the items from the protected group as a different class and then optimizing for diverse recommendations relative to this variable.

Note, however, that this type of solution does not guarantee that any given item is recommended fairly, only that recommendation lists have the requisite level of diversity. This distinction is known as list diversity vs catalog coverage in the recommendation literature and as individual vs. group fairness in fairness-aware classification~\cite{fairness}. List diversity can be achieved by recommending the same ``diverse'' items to everyone, but does not provide a fair outcome for the whole set of providers.

Achieving individual P-fairness / catalog coverage requires a more dynamic model of handling recommendation opportunities. Perhaps the closest analogy is found in on-line bidding for display advertising, where limited ad budgets serve the function of spreading impressions among competing advertisers~\cite{internetadvertisingyuan}. Consider a system in which providers have a fixed budget (in artificial system currency) that can be used to bid on recommendation opportunities during a particular time period. When a user requests a set of recommendations, appropriate recommendations are calculated, but will only be guaranteed to be delivered to the user if the advertiser successfully bids in a second-price auction mechanism for the opportunity~\cite{edelman2007internet}.

Individual P-fairness within the limitations of the personalized mechanism is achieved in this context by giving the protected group equal purchasing power to the non-protected group. Assume that there are $p$ providers from the protected group and $q$ other providers, with $p \ll q$. The purchasing parity rule requires that we allocate $B / 2p$ budget to each provider in the protected class and $B / 2q$ to the other providers, where $B$ is the total ad budget. 

Depending on the information available to each pro\-vi\-der, the design of a bidding agent for this market could be arbitrarily complex, but conversely, the system could make bidding agents simple to design by limiting the information about the recommendation opportunity that is available to the agents. A rational bidding approach would be to bid a quantity proportional to the agent's remaining budget and the candidate's quality and inversely proportional to the number of future bid opportunities. This design is similar to the BALANCE algorithm that achieves fair budget draw-down in on-line advertising~\cite{optimalmatchingKalyanasundaram}. 

\subsection{CP-fairness}

Finally, a multisided platform may require fairness be considered for both consumers and suppliers: the CP-fairness condition. This can arise in reciprocal recommendation or in any domain in which both consumers and providers may belong to protected groups: the employment scenario was discussed earlier, but there are others that we can envision. A rental property recommender may treat minority applicants as a protected class and wish to ensure that they are recommended properties similar to white renters. At the same time, the recommender may wish to treat minority landlords as a protected class and ensure that highly-qualified tenants are referred to them at the same rate as to white landlords.

The solutions described above are decoupled, in the sense that a recommendation ranking that is C-fair can be passed on to the bidding mechanism required to achieve P-fairness. One important question for future research is how the outcomes for each stakeholder and the overall system performance are affected by combining such solutions.

\section{Conclusion}

This paper extends ideas of fairness in classification to personalized recommendation. A key aspect of this extension is to note the tension between a personalized view of recommendation delivery and a regulatory view that values particular outcomes. The regulatory view is somewhat foreign to research in personalization, but there is strong argument that total obedience to user preference is not always risk-free or desirable~\cite{pariser2011filter,sunstein2009republic}. This paper also introduces the concept of multisided fairness, relevant in multisided platforms that serve a matchmaking function. Provider-side fairness, especially if defined at the individual level, requires an architecture sensitive to the dynamics of the recommendation environment.

I have outlined some possible approaches to ensuring multsided fairness in recommendation. One of the key challenges in this area is the domain-specificity of multistakeholder environments. The utilities that are delivered to each class of stakeholder are highly dependent on the type of item being recommended, the business model of the platform, and the interactions that it enables. It is therefore difficult to find appropriate data sets for experimentation and challenging to generalize across recommendation scenarios.

\section{Acknowledgments}
\noindent This work is supported in part by the National Science Foundation under grant IIS-1423368.

\bibliographystyle{ACM-Reference-Format}
\bibliography{myref.bib} 

\end{document}